\documentclass[%
 aip, numerical, 
 jmp,%
 amsmath,amssymb,
 reprint,%
]{revtex4-1}

\usepackage{graphicx}
\usepackage{dcolumn}
\usepackage{bm}
\usepackage{placeins}
\usepackage{color}

\begin{document}
\graphicspath{ {figures/}}

\title[Chaotic edge density fluctuations in the Alcator C-Mod tokamak]{Chaotic edge density fluctuations in the Alcator C-Mod tokamak}

\author{Z. Zhu}
 \email{zhuziyan@ucla.edu}
\affiliation{Department of Physics and Astronomy, University of California, Los Angeles, CA 90095, USA}
 
 \author{A. E. White}
 \affiliation{ 
MIT Plasma Science and Fusion Center, Cambridge, MA 02139, USA}
 
 \author{T. A. Carter}
 \affiliation{Department of Physics and Astronomy, University of California, Los Angeles, CA 90095, USA}
 
\author{S. G. Baek}
\author{J. L. Terry}
\affiliation{ 
MIT Plasma Science and Fusion Center, Cambridge, MA 02139, USA}

\date{\today}

\begin{abstract}
Analysis of the time series obtained with the O-Mode reflectometer
(Rhodes {\it et al} 1997 {\it Plasma Phys. and Control. Fusion} {\bf
  40} (1998) 493-510)  and the gas puff imaging (Cziegler, I. {\it et al}
2010 {\it Phys. of Plasmas} {\bf 17}, No. 5 (2010) 056120) systems
on the Alcator C-Mod tokamak reveals that the turbulent
edge density fluctuations are chaotic. Supporting evidence for this
conclusion includes: the observation of exponential power spectra
(which is associated with Lorentzian-shaped pulses in the time
series), the population of the corresponding Bandt-Pompe probability
distributions (Bandt and Pompe 2002 {\it Phys. Rev. Lett.} {\bf 88}
174102), and the location of the signal on the complexity-entropy
plane (C-H plane) (Rosso {\it et al} 2007 {\it Phys. Rev. Lett.} {\bf
  99}, 154102 (2007)).  The classification of edge turbulence as
chaotic opens the door for further work to understand the underlying
process and the impact on turbulent transport.

\end{abstract}

\keywords{turbulence, chaos, fusion, tokamaks}
\maketitle

\section{Introduction}
Turbulence transport reduces the confinement time of
magnetic-confined plasmas; understanding the nature of this turbulence
and the associated transport is therefore of great importance.  The
transport associated with plasma fluctuations, and the ability to
predict that transport, may be affected by the underlying nature of
fluctuations, that is, whether the process generating the fluctuations
is chaotic or
stochastic. Stochastic processes generate fluctuations with random
time series, and have trajectories that can sample all of phase space.  In
the presence of stochastic fluctuations, transport by random walk
diffusion is expected.  Chaotic processes, on the other hand, are
deterministic and can be generated by the interaction of a few coupled
modes (a minimum of two).  These processes live in restricted areas of
phase space (e.g. on attractors).~\cite{Origin}  As such, a random
walk diffusion model is unlikely to be a valid description of transport
arising from such a process.  In tokamak plasmas, there is a great
deal of evidence that core transport is well described by diffusive
models,~\cite{Staebler2010} but there is also evidence that the transport
may be in some instances non-diffusive.~\cite{delCastilloNegrete} Turbulence in the edge and
Scrape-Off-Layer (SOL) of tokamaks can be bursty, with intermittent
``blobs'' leading to non-diffusive transport.~\cite{Zweben2016, Terry2003,
  Carter2006}

In this work, analysis of data from two standard turbulence diagnostics, the reflectometer and the gas puff imaging (GPI)
  system, was performed, seeking to indicate the nature of edge
  turbulence in the Alcator C-Mod tokamak. The multi-channel reflectometer
  provides simultaneous localized measurement of density fluctuations at several
  radial positions covering the core to the edge of the plasma with
  excellent temporal resolution,~\cite{Reflectometry}
  whereas the GPI provides measurement of emission fluctuations within a 2D grid (radial and poloidal) that spans a $\sim 3.4 \times 3.8$ cm region near the last closed flux surface (LCFS) on the low-field-side. The spatial resolution is roughly 5 mm and the temporal 
resolution is 1 MHz. The majority of the analysis reported in this paper is on the time
series obtained by the reflectometer in the low confinement regime
(L-mode) plasmas. A comparison between reflectometry and the GPI
results and an initial analysis of fluctuations in the high
  energy confinement regime (H-mode) and improved energy confinement
  regime (I-mode) plasmas are also included. 

Three analysis techniques
are used on the time signals in order to identify whether the
underlying process is stochastic or chaotic.  These include evaluating: the
fluctuation power spectra,~\cite{Origin} the Bandt-Pompe (BP)
probability distribution,~\cite{BP2002} and the complexity-entropy
(C-H) plane.~\cite{Rosso2007} Chaotic processes have been shown to
generate power spectra that are exponential; these exponential spectra
are associated with Lorentzian pulses in the time series signal.~\cite{Origin}
The BP probability distribution provides information on the structure
of the time series signal by evaluating the distribution of amplitude
orderings in the signal.  The C-H plane analysis makes use of the BP
probability distribution to classify the nature of the signal based on
the so-called complexity and the BP entropy. The BP probability
distribution and the C-H plane analysis have been successfully applied to analyze a
number of time series including financial data,~\cite{Zanin2008, Zunino2009} mammal neural activities,~\cite{ouyang2009deterministic} edge density fluctuations in tokamaks,\cite{Maggs2015, Maggs2013}, and MHD turbulence. \cite{weck2015permutation}
The results of the application of these three techniques indicate that the edge density fluctuations in
L-mode, H-mode and I-mode plasmas in the Alcator C-Mod tokamak are
chaotic. Similar chaotic edge density fluctuations were observed in a
wide range of plasma devices of different geometries, such as in the
DIII-D tokamak,~\cite{Maggs2015} the TJ-K stellarator~\cite{Hornung2011} and the Large Plasma Device.~\cite{PacePRL, Maggs2011}

This paper is organized as follows: Sec.~\ref{cond} briefly introduces
the experimental device and discusses the two
diagnostics (the O-mode reflectometery and the GPI) used
in the experiments. Section~\ref{techniques} introduces the theoretical
background of this study and the three main techniques used to
distinguish between chaotic and stochastic
signals. Section~\ref{analysis} presents analysis of the experimental data using the techniques introduced in section~\ref{techniques}. Section~\ref{dis} discusses the findings and presents conclusions.

\section{Experimental Setup}\label{cond}
The experiments presented here were performed in the compact, high-field Alcator C-Mod tokamak ~\cite{Marmar2009}. The device has major radius $R =
0.67$m, minor radius $a = 0.22$m. In this work, a total of 61 shots
are analyzed. A wide range of parameters and conditions are included
in these shots: the toroidal magnetic field $B_\mathrm{T}$ ranges from
2.7 to 8 T; plasma current $I_\mathrm{p}$ ranges from 0.5 to 1.2 MA;
the line averaged density ranges from $0.5$ to $1\cdot10^{20}
\mathrm{m^{-3}}$; the edge safety factor q95 ranges from 3 to
7; and the RF heating power $P_{\mathrm{RF}}$ ranges from 0.6 to 4.5 MW. Different confinement regimes were studied: low-energy confinement
regime (L-mode), the high energy confinement regime (H-mode) and
improved-energy confinement regime (I-mode). Some Ohmic plasmas (without RF heating) are also included. In each shot, 20 ms temporal signals were chosen based on the radial density profile measured by Thomson scattering diagnostics,~\cite{Hughes2001} with the assumption that there is no significant change in the density profile within 20 ms. 

The first diagnostic discussed in this work is the multi-channel O-Mode reflectometer.~\cite{dominguez2012study} Signals $S(t)$ collected by the reflectometer are complex and are composed of an amplitude component $E(t)$ and a phase component
$\phi(t)$, $S(t) = E(t)\ e^{i\phi(t)}$. The real part of a signal is referred as
the inphase, $S_{\mathrm{re}} =
\mathrm{Re}[E(t)\ e^{i\phi(t)}]$, and the imaginary
part is referred as the quadrature, $S_{\mathrm{im}} =
\mathrm{Im}[E(t)\ e^{i\phi(t)}]$.\cite{Reflectometry}
A total of 5 reflectometer channels was analyzed, giving measurements
at a range of radial positions. The frequencies of the 5 channels are: 50 GHz, 60 GHz, 75 GHz, 88 GHz, and 112 GHz, with the cutoff density ranging from $0.3 - 1.5 \cdot 10^{20} \mathrm{cm^{-3}}$. Advantages of reflectometry include its high temporal resolution and capability of localized measurements. However, it is difficult to extract absolute fluctuation levels from reflectometer signals and the radial position of the measurement is not fixed in space, but varies with plasma density. The measurement position can be tracked during a shot by comparing the cutoff frequency with the density profile measured by a Thomson scattering diagnostic.\cite{Hughes2001} 

The gas puff imaging (GPI) was also employed to measure turbulent edge
fluctuations.~\cite{Terry2003, Zweben2016, GPI2} Helium gas is puffed locally into the Scrape-Off Layer, and HeI line emission (587.6 nm) is monitored along sightlines that cross the puff region toroidally. Since the line emission is due to electron-impact excitation by the local plasma, the emissivity is a function of both $n_e$ and $T_e$ and responds to fluctuations in those plasma quantities. The emission rate can be parameterized as ${\mathrm S}\ (\mathrm{photon\cdot s^{-1} \cdot cm^{-3}})$ $\propto (n_e)^{\alpha} \cdot (T_e)^{\beta}$, where $\alpha$ and $\beta$ depend on the time-averaged local quantities $n_e$ and $T_e$.\cite{GPI}. 
The GPI provides a 2D image of the normalized emission in the radial
and poloidal directions.~\cite{GPI} Unlike reflectometry, the exact
spatial positions of GPI-measured fluctuations are fixed by the
viewing optics. In this study, GPI signals were located at radial
positions ranging from roughly 1.5 cm inside the LCFS to 1.5 cm outside the LCFS at single height that is 2.4 cm below the outboard midplane. The reflectometry and GPI measurement were separated by ~18 degrees toroidally. When we compare the results from the two, we do so by mapping them using EFIT Equilibrium and Reconstruction Fitting Code~\cite{lao1985reconstruction} to the same flux surface and use signals taken during the same time period. Both
diagnostics were sampled at 2MS/s.

\section{Identifying chaotic vs. stochastic signals}\label{techniques}
Although chaotic and stochastic signals have distinct origins, they
can be difficult to distinguish due to their similarities: both will
give rise to time series that have broadband power spectra and seemingly random behavior.\cite{Rosso2007} This section introduces the three analysis tools to distinguish time series generated by chaotic and stochastic processes: (1) the shape of the power spectra,  (2) the population of Bandt-Pompe (BP) probability distributions, and (3) the corresponding complexity-entropy (C-H) plane.

\subsection{Power spectra and corresponding time series}
It has been established by researchers in different disciplines since the 1980s that an intrinsic characteristic of deterministic chaos is the exponential power spectrum; i.e: $P(\omega) \propto \mathrm{exp}(-2\omega\tau)$, whereas stochastic processes are associated with power law spectra.\cite{Frisch1981, Greenside1982, Libchaber1983} Although both chaotic and stochastic processes have broadband power spectra, exponential chaotic spectra can be formed by a small number of coupled modes. Such exponential power spectra correspond to Lorentzian shaped pulses in the time signals (eqn. \ref{eqn:lorentz}). 

\begin{equation}
\label{eqn:lorentz}
L(t) = \frac{A}{\tau^2+(t-t_0)^2}
\end{equation}

In Eqn.~\ref{eqn:lorentz}, $\tau$ is referred to as the pulse width,
which is the half width at half maximum; $t_0$ is the center of the
pulse, and $A$ is a normalization constant. Lorentzian-shaped pulses
in time are produced by particle motions in the vicinity of the
separatrix boundaries of elliptical regions in flow fields, or more
generally, near the limit cycles of attractors in nonlinear dynamical
models.~\cite{Origin} The pulse width is determined by the
imaginary part of the eigenvalues of the Jacobian of the flow fields,
which is essentially the angular frequency of each trajectory around
the attractor. In general, the flow fields are associated with Lorentzian
bifurcation transports scalar quantities, and in two-dimensional
bifurcation, if a scalar quantity has a linear gradient in the $y$-direction
(such as density and temperature gradients in magnetic-confined fusion
devices), a Lorentzian shape in the $y$ component of the potential field leads to the Lorentzian-shaped pulses in the trajectories.~\cite{Origin}

The power spectrum for a series of $n$ Lorentzian pulses is then as follows,

\begin{equation}
\label{eqn:power_n}
\tilde{P}(\omega) \propto \sum_{n}{e^{-2 \omega \tau_n}}
\end{equation}

Therefore, if a temporal signal has a series of Lorentzian pulses with
a well-defined pulse width, one should expect an exponential power
spectrum, which exhibits a linear shape on a semi-log scale. In this
way, a broadband power spectrum can be formed by a low-dimensional
chaotic process where a single time scale defines the dynamics.  This
is possible with a minimum of two interacting modes (the trajectory of
particles in the potential fields of the two modes is
chaotic).~\cite{Origin}
The slope of a Lorentzian power spectrum on a semi-log scale is $-2\omega \tau = - 4\pi \tau f$ (with the assumption of a single pulse width in the time history). Then, the value of $\tau$ can be found by fitting the slope of the spectrum on the semi-log scale:

\begin{equation}
\label{tau_fit}
\tau = -\frac{\mathrm{slope}}{4\pi}
\end{equation}

A complex time signal is constructed from the measurements of the
amplitude and the phase of the reflected signal associated with the
reflectometer diagnostic.   The Fourier transform of a complex time
signal has both negative and positive frequencies. In this work, the slope is fitted only for the positive
frequencies since all presented spectra are roughly symmetric about the zero frequency. However, the GPI power spectra presented only have positive
frequencies because GPI signals measure a single quantity (light
fluctuation amplitude) and do not contain phase information, and thus were
compared with real amplitude spectra from the reflectometer.

\subsection{Bandt-Pompe probability and complexity-entropy plane}
\FloatBarrier

\begin{figure}[h]
\centering
\includegraphics[width=\linewidth]{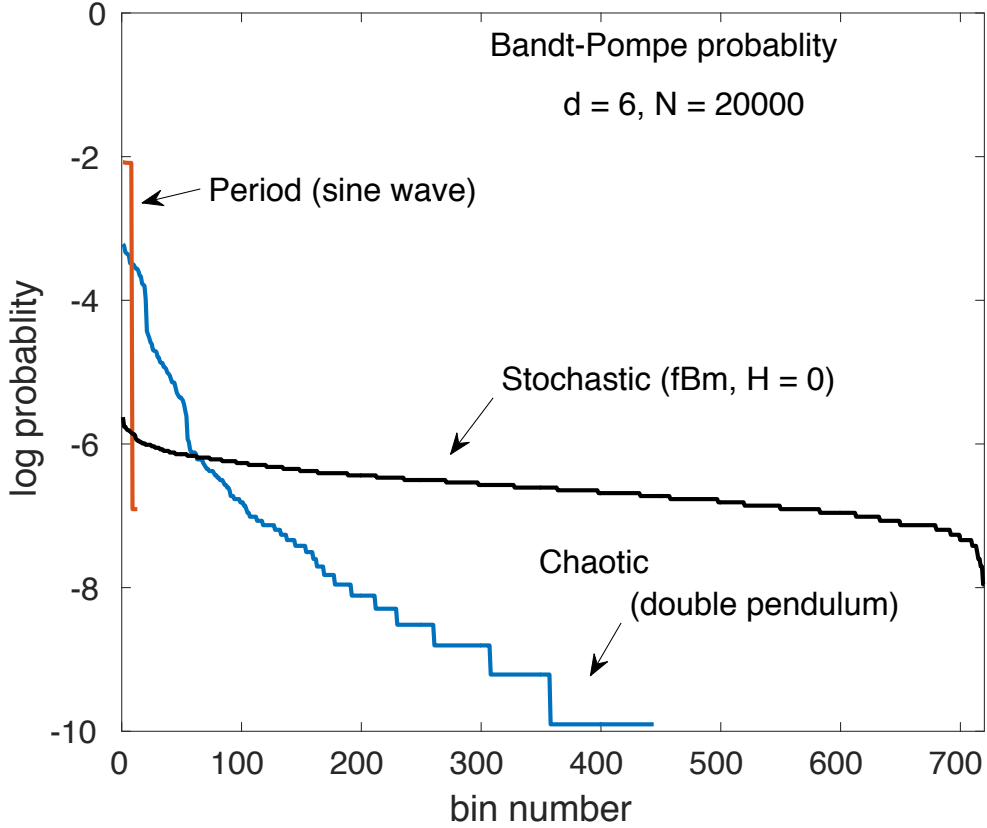}
\caption{The BP probability distributions of stochastic, chaotic, and periodic test data with $N = 20, 000$ and embedding dimension $d = 6$. The stochastic signal is the fractional Brownian motion with Hurst exponent $H_{\mathrm{exp}} = 0$, \cite{Mandelbrot1968} and it is uniformly distributed over the BP probability space. The chaotic signal is the trajectory of a ball in a double pendulum system\cite{marion_thornton_1995}, and it has a range of highly-occupied states and a range of unoccupied states. The periodic test signal is a sine wave $f(t) = \mathrm{sin} (\omega t)$, where $\omega = (2 \pi \cdot 500) \ \mathrm{rad/s}$, and it only has very narrow range of occupied states.} 
\label{fig: BP_test}
\end{figure}

\begin{figure}[h]
\centering
\includegraphics[width=\linewidth]{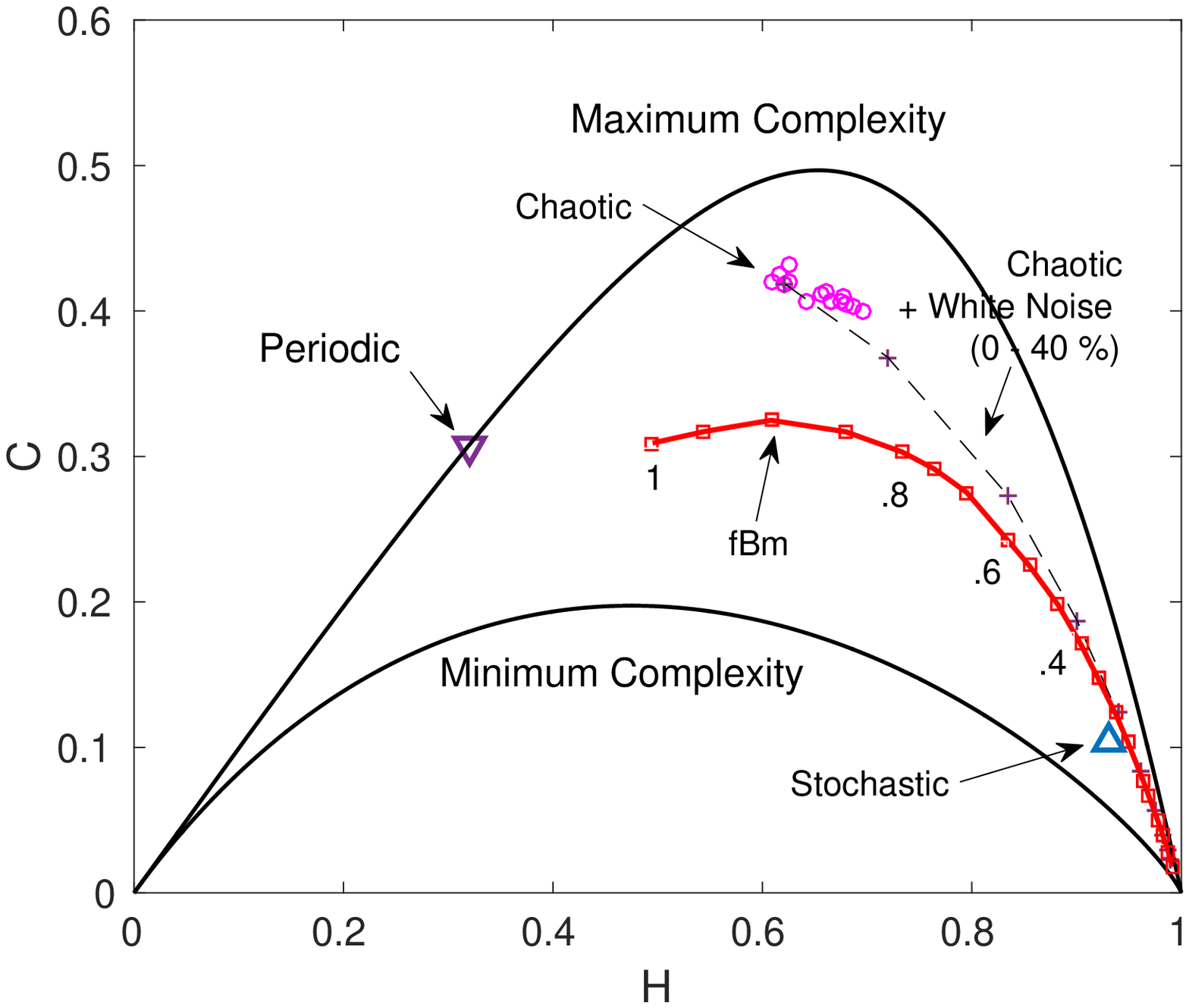}
\caption{Periodic, chaotic, and stochastic signals fall on different parts of the C-H plane. The red curve shows fractional Brownian motions with different Hurst exponents (marked below the red curve) $0 \leq H_{\mathrm{exp}} \leq 1$, \cite{Mandelbrot1968} which is purely stochastic. The second set of stochastic data is generated using randomly distributed Lorentzian pulses with different pulse widths (marked with blue triangle), which will be discussed later in section \ref{dis}. The periodic test signal is a sine wave $f(t) = \mathrm{sin} (\omega t)$, where $\omega = (2 \pi \cdot 500) \ \mathrm{rad/s}$; it has relatively low $C$ and $H$. The chaotic data was generated using the trajectories in a double pendulum system with different initial conditions;\cite{marion_thornton_1995} the points are located in between the maximum complexity curve and the fractional Brownian motion curve. The dashed trajectory was corresponds to adding 0 - 40\% white noise to one of the chaotic time traces. $H$ increases as the percentage of white noise increases. }
\label{fig:CH_test}
\end{figure}

Stochastic and chaotic processes are distinct in the phase space that the time
signals have access to and therefore in the structure of the generated
time signals. Two techniques to distinguish the structures generated by
chaotic and stochastic processes are the Bandt-Pompe (BP) probability
distribution and the complexity-entropy (C-H) plane. The BP
probability quantifies the frequency of occurrence of structure in
time signals by using permutations of the ordering of the amplitudes
of consecutive values, at evenly separated discrete
points.~\cite{BP2002} To compute
the BP probability, an ``embedding space'' with dimension $d$ is used. So-called ``d-tuples'' are generated by taking $d$ consecutive
points in the time signal and ordering them from largest amplitude to
smallest.  For example at time points $t = (51, 52, 53, 54)$, the signal
takes on values $y = (5, 9, 7, 10)$.  The 4-tuple created by these points
would be $(4, 2, 3, 1)$, indicating the largest amplitude occurs in the
4th time point, followed by the 2nd time point, and so on.   In a time series with $N$
points, there is a total of $(N - d + 1)$ d-tuples: $(x_1, x_2, ...,
x_{N-d+1})$. For each embedding dimension $d$, there are $d!$ possible
permutations, which is the number of possible amplitude ordering
combinations. From the d-tuples, one could compute the relative
frequency or the probability for a given amplitude ordering
permutation type $\pi$:\cite{BP2002}

\begin{equation}
\label{relative_freq}
p (\pi) = \frac{\#\{t | t\leq N - d + 1, (x_t)\ \mathrm{has\ type\ } \pi\}}{N - d + 1} 
\end{equation}

The BP probability space thus has dimension of $d!$, and is normalized
to 1. The way the BP probability distributions are presented is that
one ranks the relative frequencies of each permutation from the
highest to lowest and plot the relative frequencies against bin number
(permutation number of the amplitude ordering) using a semi-log
scale. If the relative frequency of a permutation is zero, that bin is
empty on presented BP probability plots. In this way, the BP
probability distribution is capable of detecting the preferable
population in the temporal signal and thus the structure of amplitude
orderings.

The choice of $d$ depends on the value of $N$ and the time scale
$d\Delta t$ of the structure being investigated, where $\Delta t$ is the time interval between adjacent data points.~\cite{Maggs2015} On
one hand, one should choose a small $d$ such that $N/d! \gg 1$ to
obtain a reliable result. On the other hand, $d$ cannot be so small
that relevant structures in the signal (e.g. Lorentzian pulses) cannot
be represented appropriately. A typical choice of $d$ is usually
within the range of $3 \leq d \leq 7$. In this work, $N = 20,000$ and
$d = 6$ were used for all the BP probability and C-H plane
analysis. With $N/d! \approx 27.3$, the number of d-tuples is large enough to capture the structure. With $d\Delta t = 1.2 \mathrm{\mu s}$, each d-tuple has a time scale that is comparable to the typical pulse width. Therefore, the resulting BP probability is able to capture relevant structures in the signal. Analyses using $d = 5$ and $d = 7$ were compared, but the qualitative results are the same. A more detailed study about how to choose can be found in the Ph.D thesis of B. Frettman. \cite{friedman2013simulation}

In general, a stochastic signal has a relatively uniform BP
distribution because it does not have a preferable amplitude ordering
and thus occupies all the BP probability space equally likely. A
chaotic signal would have a peak around small bin number because it
contains structures and thus has preferable states in the BP
probability space: the distribution has a range of states with high
relative frequencies along with a wide range of unoccupied or
low-occupied states. A periodic signal has repeating structures and
thus has very few occupied states and is very sharply peaked in the
probability space. In Fig.~\ref{fig: BP_test}, BP distributions
generated from three groups of test signals
(stochastic, chaotic, and periodic) are shown.

The shape of BP probability distributions provides a qualitative view
of the nature of a given time history, and it can be quantified using
the complexity-entropy plane (C-H plane). This technique was
introduced by Rosso, {\it et al}~\cite{Rosso2007} to distinguish
chaotic signals from noise. The C-H plane consists of two statistical
measurements: the normalized Shannon entropy $H$ and the
Jensen-Shannon complexity $C$. The Shannon entropy measures the
disorder or the uncertainty of a physical process described by a
probability distribution. For example, for a given probability
distribution, if the outcome of a physical process is random or
unpredictable, the process has high entropy. Statistical complexity
measures the structure of a time signal in terms of the uncertainty of
a system and its ``distance'' from complete
disequilibrium.\cite{Martin2006}

First, the Shannon entropy $S$ for a given probability distribution $P
= (p_1, p_2, ..., p_{d!})$ with d! possible states is defined as:

\begin{equation}
S(P) = -\sum_{i = 1}^{d!} p_i \log_2{p_i}
\label{eqn: S}
\end{equation}

The normalized Shannon entropy $H$ is then computed by dividing $S$ by
the maximum entropy, which corresponds to uniform probability
distribution $P_e$, with $p_i = 1/d! \ \forall \ i \in [1, d!]
$. $S_{\mathrm{max}} = \log_2{d!}$. The uniform distribution
corresponds to the most random/unpredictable system, and thus has
maximal entropy.

\begin{equation}
\label{eqn:H}
H(P) = \frac{S(P)}{S_{\mathrm{max}}}=-\frac{1}{\log_2{d!}}\ \sum_{i = 1}^{d!} p_i \log_2{p_i}
\end{equation}
where $0 \leq H(P) \leq 1$. 

The statistical complexity of a given probability distribution $P$ is
defined as the product of entropy and disequilibrium, the ``distance''
between $P$ and $P_e$.\cite{Martin2006} There are different
definitions of complexity depending on how disequilibrium is
defined. In {\it{Rosso, et al}},\cite{Rosso2007} the disequilibrium
$Q_J (P)$ is defined in terms of the Jensen-Shannon divergence
$D_{JS}$, given as:

\begin{eqnarray}
Q_J (P) = Q_0 \cdot D_{JS}=\nonumber\\
 Q_0\left[S(\frac{P + P_e}{2}) - \frac{S(P)}{2}
 - \frac{S(Pe)}{2}\right]
\label{eqn:disequilibrium}
\end{eqnarray}
where the notation $\frac{P+P_e}{2}$ denotes adding the BP probability
$p_i$ to the uniform probability $p_e = 1/d!$ and then dividing by 2
$\forall \ i \in [1, d!]$.  $Q_0$ is a normalized constant $Q_0 = -2{\left[\frac{d!+1}{d!}\log_2{(d!+1)} - 2\log_2{(2d!)}+\log_2{d!}\right]}^{-1} $ such that $0 \leq Q_J (P) \leq 1$. 

The complexity of a BP probability $P$ is then defined as, 

\begin{equation}
C(P) = Q_J (P)\cdot H(P)
\end{equation}

\begin{equation}
C(P) = -2\frac{S(\frac{P + P_e}{2}) - \frac{S(P)}{2} - \frac{S(Pe)}{2}}{\frac{d!+1}{d!}\log_2{(d!+1)} - 2\log_2{(2d!)}+\log_2{d!}} H(P)
\label{eqn:complexity}
\end{equation}

For each given BP probability distribution, there is a corresponding
$C$ and $H$ value and thus a point on the C-H plane. Since BP
probabilities with the same entropy do not necessarily have the same
complexity, the $C$ and $H$ can be used as independent parameters for
the investigated signals when plotting the C-H
plane.\cite{Maggs2015} For a given $H$, there is a maximum and a
minimum possible complexity values, giving a complexity boundary
for $0 \leq H \leq 1$. All points should fall between the two
curves. {\it Calbet} and {\it L\'opez-Ruiz} ~\cite{Calbet2001} and {\it Martin, et
  al.}~\cite{Martin2006} discussed the methods to solve for the extrema
of different definitions of complexity using Lagrangian multipliers in
detail. For the Jensen-Shannon complexity, the probability
distributions that minimize and maximize the complexity are given in
Table~\ref{Cmin} and Table~\ref{Cmax} respectively. The boundaries are different
for different choice of $d$.

\begingroup
\squeezetable
\begin{table}[h]
\begin{ruledtabular}
\caption{\label{Cmin} The probability distributions that minimize the complexity.}
\begin{center}
\begin{tabular}{  c  c  c  }

Number of states with $f_j$ &   $\ \ \ f_j \ \ \ $   & Range of $f_j$\\

1 & $f_{\mathrm{min}}$ & $\left[\frac{1}{d!}, 1\right] $ \\
$d! -1$&$\displaystyle\frac{1-f_{\rm{min}}}{1-d!}$&$\left[0, \frac{1}{d!}\right]$\\

\end{tabular}
\end{center}
\end{ruledtabular}
\end{table}
\endgroup

\begingroup
\squeezetable
\begin{table}[h]
\begin{ruledtabular}
\caption{\label{Cmax} The probability distributions that maximize the complexity, where $n \in \mathbb{Z}$ and $0\leq n \leq (d! -  1)$; the maximum complexity curve is not smooth}
\begin{center}
\begin{tabular}{c  c c }

Number of states with $f_j$ &   $\ \ \ f_j \ \ \ $   & Range of $f_j$\\

$n$ &0&0\\
1 & $ f_{\mathrm{max}}$ & $\left[0, \frac{1}{d! - n}\right]$\\
$d!-n-1$ & $\displaystyle\frac{1-f_{\mathrm{max}}}{d!-n-1}$ & $\left[\displaystyle\frac{1}{d!-n-1},\ \displaystyle\frac{1}{d!-n-1}\right]$\\
\end{tabular}
\end{center}
\end{ruledtabular}
\end{table}
\endgroup

Different locations on the C-H plane correspond to different types of
processes generating the corresponding temporal
signals. Figure~\ref{fig:CH_test} shows what different regions on the
C-H plane represent: complexities and entropies of double pendulum
trajectories (chaotic),\cite{marion_thornton_1995} chaotic signals with different percentages of
added white noise (transitioning from dominantly chaotic to dominantly
stochastic),\cite{Rosso2012} and a sine wave (periodic). Generally speaking, periodic
systems have low entropies and complexities because of their
repetitive patterns and predictability. Chaotic systems have
high complexities and medium-ranged entropies.  Stochastic systems
have high entropies and low complexities because they are highly
uncertain and they are close to the uniform distribution. While there
is no hard boundary between chaotic and stochastic signals, the
fractional Brownian motion (fBm) \cite{Mandelbrot1968} is a useful curve to compare the data
points with: the closer a point is to the fBm line from above, the
more stochastic it is. Points on the fBm line or below are considered purely
stochastic. 

\begin{figure}[h]
\centering
\includegraphics[width = 0.9\linewidth]{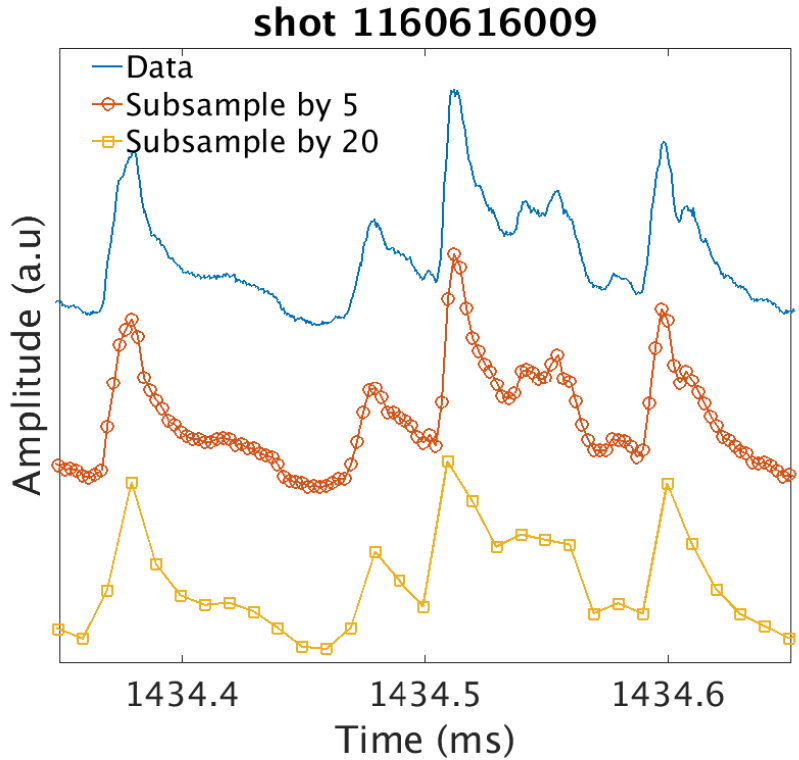}
\caption{Turbulent fluctuations obtained by the gas puff imaging (GPI) with different subsampling rate $r_\mathrm{s}$ show the effect of subsampling on temporal signals. The first signal (blue) is with $r_\mathrm{s} = 1$ (no subsampling); high frequency noise is observed. The second signal (red) is with $r_\mathrm{s} = 5$. At this subsampling rate, the original structure of the signal is retained while getting rid of the high frequency noise. The third signal (yellow) shows $r_\mathrm{s} = 20$. This sampling rate fails to capture the true nature of this signal. }
\label{subsample_time}
\end{figure}

\begin{figure}[h]
\centering
\includegraphics[width=\linewidth]{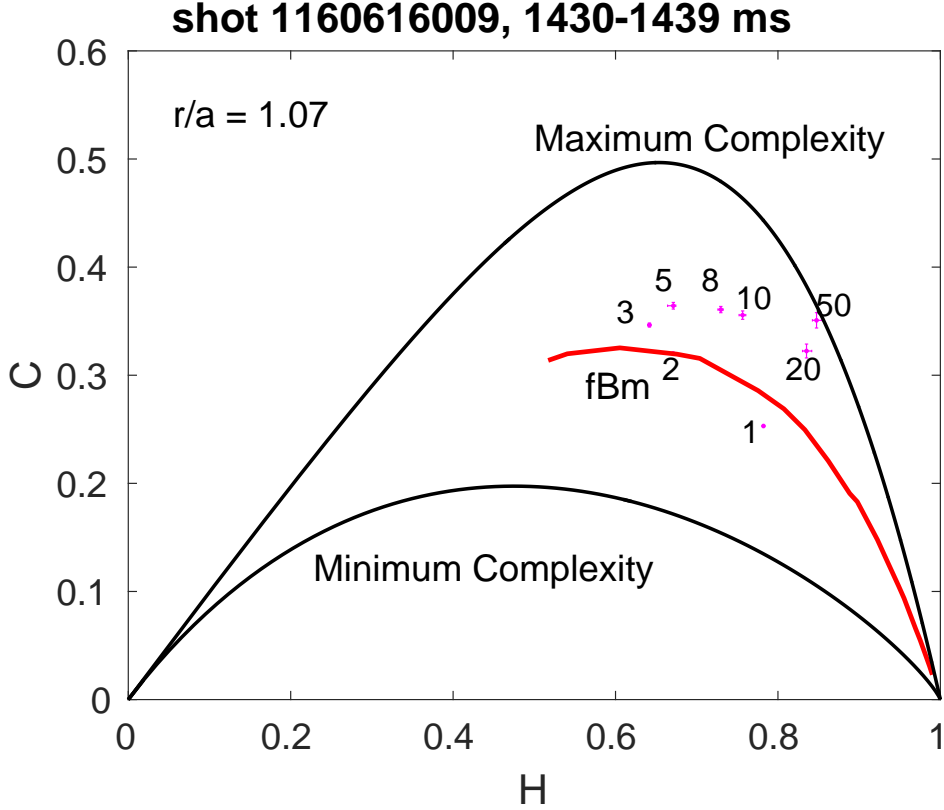}
\caption{The GPI data with different subsampling rate $r_\mathrm{s}$'s on the C-H plane shows that $r_\mathrm{s}= 5$ is the best subsampling rate to apply. With $r_\mathrm{s} = 1$, the data point is below the fBm line. As $r_s$ increases to $r_\mathrm{s} = 5$, points move towards the chaotic region. As $r_\mathrm{s}$ keeps increasing, points move back towards the stochastic region, meaning the signal is over-subsampled. Notice that when $r_\mathrm{s} = 50$, the point falls close to the $C_{\mathrm{max}}$ curve; this is because when the sampling rate is too big, $N/r_\mathrm{s}$ is small such that there are only a few occupied states in the BP probability distributions, which assemble the distributions that maximize the complexity. }
\label{subsample_CH}
\end{figure}

\subsection{Subsampling}
Subsampling is a technique being applied to the C-H analysis on
signals with high frequency noise.\cite{Maggs2013} Although
subsampling is not necessary for refletometry analysis, it is required
for the GPI analysis because the signal is dominated by white noise
above 300 kHz (see fig. \ref{spectra_GPI} for example).

When calculating the BP probability, if $d\Delta t$ is small compared
to the time scales of interest (e.g. structures in the signal), each
d-tuple is strongly influenced by the high frequency noise rather than
the actual structure of the signal being investigated. Therefore,
subsampling is required to ensure that the set of d-tuples reflects
the actual structure in the data. To compute the BP probability, one
should choose a set of N equally-spaced points. To perform subsampling
on the dataset is to increase the spacing between adjacent time points
by a factor of the subsampling rate, denoted as $r_\mathrm{s}$, which
is essentially to decrease the Nyquist frequency by a factor of
$r_\mathrm{s}$. In this way, $r_\mathrm{s}$ sets of data points with
size $N/r_\mathrm{s}$ are obtained, which correspond to a number of
$r_\mathrm{s}$ different BP probability distributions and points on
the C-H plane (fig. \ref{subsample_CH}). Errorbars can be obtained by
calculating the standard deviation of the set of $C$s and $H$s
corresponding to a given dataset. The choice of $r_\mathrm{s}$ is not
arbitrary but rather restricted. It is important to choose a best
$r_\mathrm{s}$ that is not only able to remove the noise component but
also able to resolve the actual nature of the signal
(fig. \ref{subsample_time}). Another problem with a large
$r_\mathrm{s}$ is that $N/(r_\mathrm{s}\cdot d!)$ might be too small
such that the BP probability distribution is not reliable even if the
structure in the temporal signals could be captured.

\section{Experimental Signal Analysis} \label{analysis}

\begin{figure}[h]
\centering
\includegraphics[width=\linewidth]{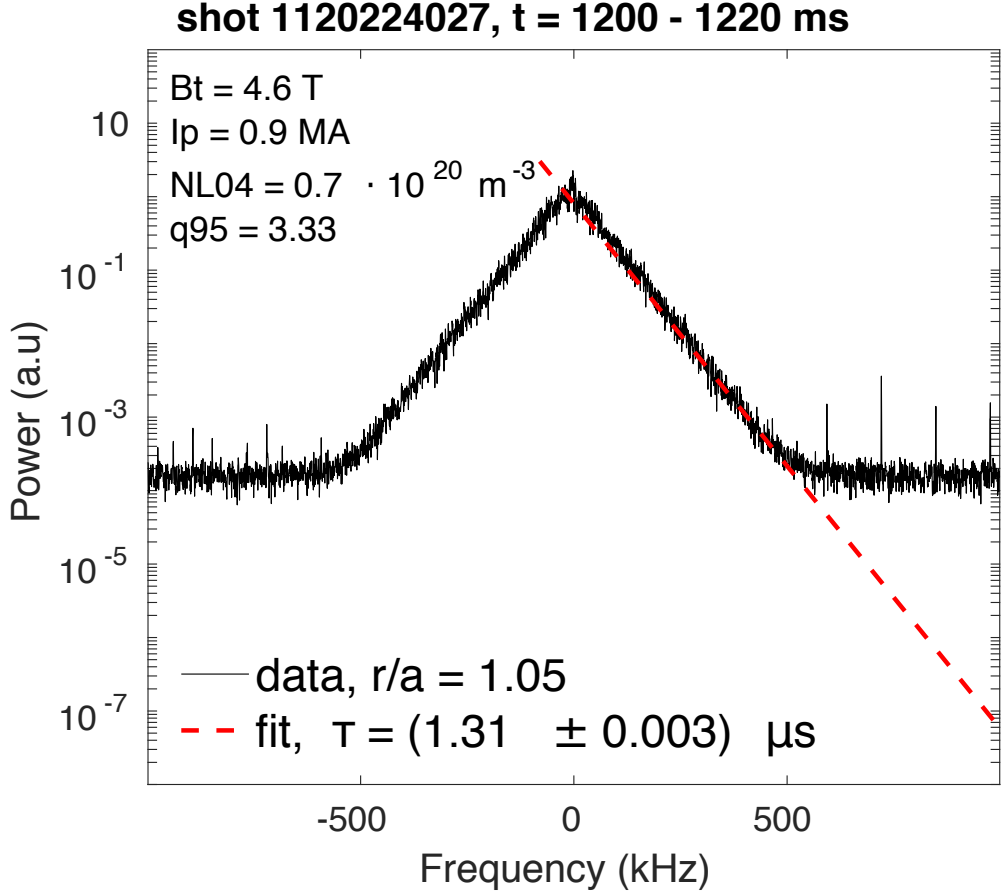}
\caption{A typical power spectrum of edge density fluctuations from L-mode plasmas as measured by reflectometry exhibits clear exponential shape; fitting the spectra to $P(\omega) \propto \exp{(-2\omega\tau)}$ gives the pulse width $\tau$. The power spectra are normalized.}
\label{spectra}
\end{figure}

This section utilizes the three techniques introduced in Section III to address the question of whether the turbulent fluctuations in the Alcator C-Mod tokamak are generated by chaotic or stochastic
processes. As noted previously, the edge fluctuations being analyzed are those from the O-mode reflectometry and the GPI. L-mode plasmas were analyzed in detail and a brief
overview of H-mode and I-mode signals is also included. In all cases: (1) the edge
power spectra exhibit a clear exponential shape $P(\omega) \propto
\exp{(-2\omega\tau)}$; (2) the BP probability distributions show a range
a concentration of probability in a subset of permutations; and (3)
the generated points on
the C-H plane fall on the chaotic regions.  These three observations
are a strong indication that the process generating edge density
fluctuations in Alcator C-Mod is chaotic in nature.

\subsection{Exponential power spectra}

\begin{figure}[h]
\centering
\includegraphics[width = \linewidth]{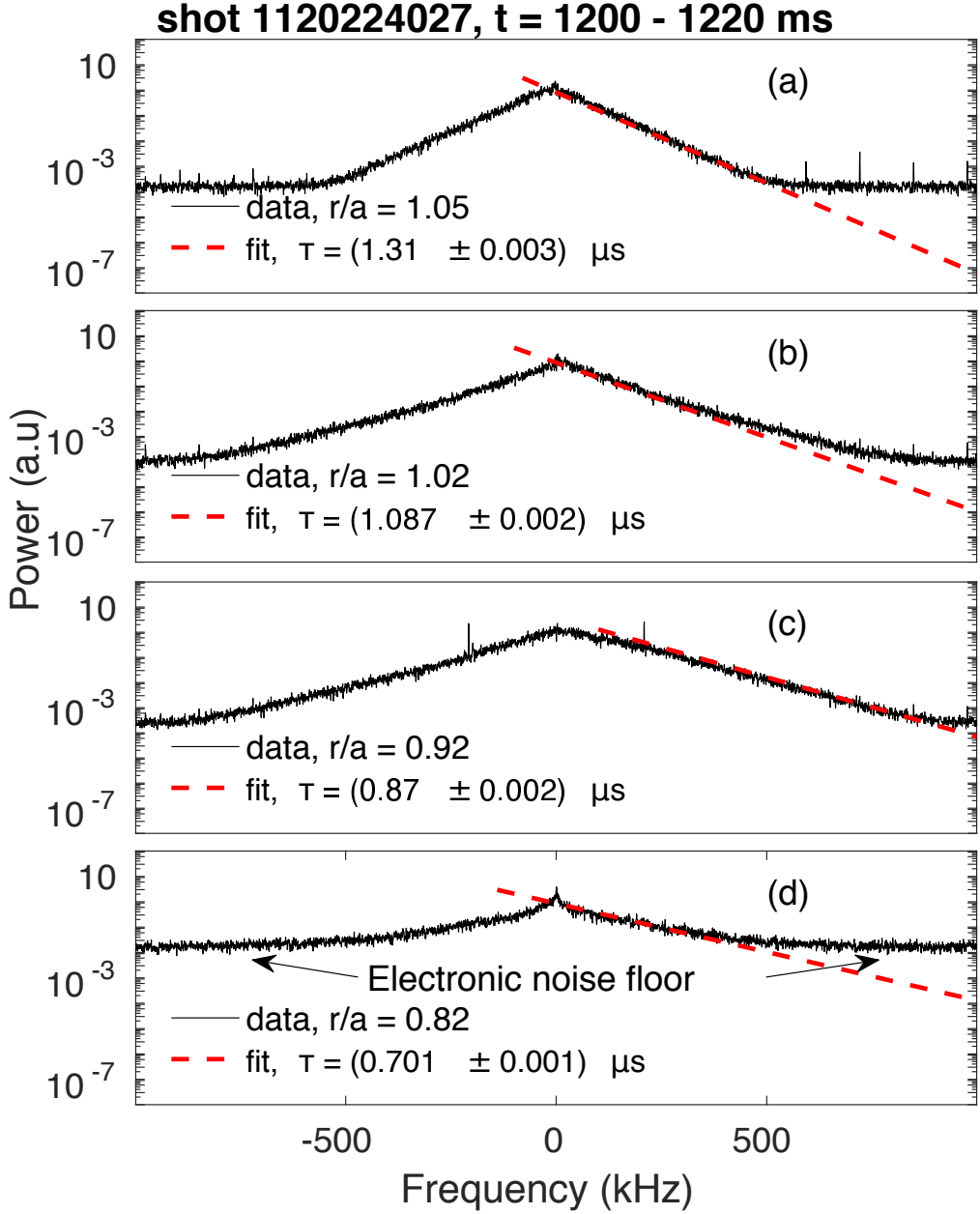}
\caption{Normalized power spectra from four different reflectometer channels (different major radii in the machine) shows a trend that $\tau$ decreases as the major radii decreases.}
\label{spectra_R}
\end{figure}

As mentioned earlier, reflectometer data from each channel is complex,
consisting of a inphase (real) part and a quadrature (imaginary)
part. Therefore, the power spectra presented have both positive and
negative frequencies that are not necessarily symmetric around the zero
frequency. Figure~\ref{spectra} shows power spectra for
reflectometer signals inside and outside of the last closed flux surface. These power spectra were
fitted to the following equation:

\begin{equation}
\tilde{P}(f) = A\cdot e^{-4 \pi \tau f  } + \mathrm{const.}
\end{equation}

where $\tau$ is the pulse width of the corresponding Lorentzian pulses in time.

To obtain the best fit and the error, a technique called maximum
likelihood estimation (MLE) was used.~\cite{Newman2005} It is a
technique to optimize the parameter that fits data to a model by
maximizing the likelihood function of the model. The error can be
obtained by calculating the standard deviation or the square root of
the variance. For an exponential fit, we use the following normalized
model and treat it as a probability distribution:

\begin{equation}
p(f) = A\cdot e^{-bf} = b\cdot e^{-b(f-f_{\rm{min}})}
\end{equation}
where $b = 4\pi\tau $, $f \geq f_{\rm{min}}$ and $f_{\rm{min}} > 0$ for $p (f)$ to be normalizable.

Given a frequency-dependent power spectrum $P$, we use it as a
probability distribution with frequency dependence, and generate a set
of $n$ values of ${\mathbf f} = \{{f_i} \}$ based on the
distribution. The probability that the data is generated from this
distribution is proportional to the log-likelihood of the dataset
$\mathbb{L}$, defined as follows,

\begin{equation}
\mathbb{L} = \ln{P({\bf{f}} | b)} = n \ln{b} + nbf_{\rm{min}} - b \sum_{i = 1}^{b} f_{i}
\end{equation}

Maximizing $\mathbb{L}$ by taking the derivative of $\mathbb{L}$ with
respect to $b$, setting it to zero and then solving for $b$:

\begin{equation}
b = \frac{1}{\frac{1}{n}(\sum_{i=1}^{n} \ f_{i}) - f_{\rm{min}}}
\end{equation}

The error of the optimal parameter can be obtained by first taking the
exponential of the log-likelihood function (which is the likelihood
function) and then calculating the half width of the maximum
likelihood function, i.e: the standard deviation $\sigma_{b}$,

\begin{equation}
\exp{(\mathbb{L})} = b^n \exp{\left[b\left(n f_{\mathrm{min}} - \sum_{i = 1}^{n} f_i\right)\right]}
\end{equation}
Define $ a = nf_{\mathrm{min}} - \sum_{i = 1}^{n} f_i$.

\begin{equation}
<b^2> = \frac{\int_{0}^{\infty} db\ b^{n+2} e^{ab}}{\int_{0}^{\infty}db\ b^{n} e^{ab} } = \frac{(n+1)(n+2)}{a^2}
\end{equation}

\begin{equation}
<b> = \frac{\int_{0}^{\infty} db\ b^{n+1} e^{ab}}{\int_{0}^{\infty}db\ b^{n} e^{ab} } = \frac{n+1}{a}
\end{equation}

\begin{equation}
\sigma_{b} =  \sqrt{<b^2> - <b>^2} = \frac{\sqrt{n+1}}{\sum_{i=1}^{n}f_i - nf_{\rm{min}}}
\label{error_b}
\end{equation}
which gives the error of the slope. The uncertainty on the evaluation of  $\tau$, $\sigma_{\tau} = -\sigma_b/{4\pi}$.

Historically, power spectra of turbulent fluctuations have been
presented in log-log scale because the Kolmogorov $5/3$ law of
turbulence \cite{Kolmogorov1941} states that the turbulent power
spectra should fit to power laws. A quantitative assessment on whether
a power law model or an exponential model provides a better fit to the
edge turbulence data is Akaike information criterion
(AIC).~\cite{Newman2005,Edwards2007} The AIC is a measure of the
relative quality of statistical models for a given set of data. For a
given dataset $\mathbf{f_i}$, the corresponding AIC is defined as,

\begin{equation}
\mathrm{AIC}_i = -2 \ln{\mathbb{L}_i(\hat{\theta_i}|{\mathbf{f}}}) + 2K_i
\end{equation}
where $\mathbb{L}_i$ is the log-likelihood function for a given
model. $\hat{\theta_i}$ is the most probable parameter; $K_i$ is some
constant, and is the same for power law and exponential models.

The Akaike weights are the relative likelihood of each model (normalized to 1):

\begin{equation}
w_i = \frac{e^{-\Delta_i/2}}{e^{-\Delta_1/2}+e^{-\Delta_2/2}}
\end{equation}
where $\Delta_{i} = \mathrm{AIC}_i - \rm{AIC_{\rm{min}}}$.

For the sake of completeness, the log-likelihood of a normalized power law model $p(f)  = \frac{\alpha - 1}{f_{\rm{min}}} \cdot \left(\frac{f}{f_{\rm{min}}}\right)^{-\alpha}$ is given as follows,

\begin{equation}
\mathbb{L} = \ln{P({\mathbf{f}} | \alpha)} = \sum_{i = 1}^{n} \ \left[\ln{(\alpha-1)} - \ln{f_{\rm{min}}} - \alpha \ln{\frac{f_{i}}{f_{\rm{min}}}}\right]
\end{equation}

In all presented cases, the Akaike weight of the exponential model is
1 and that of the power law model is 0, which quantitatively confirms
that the exponential model is a better fit than the power law fit.

Power spectra at different major radii are shown in
Fig.~\ref{spectra_R}. Different reflectometer channels send out
electromagnetic waves at different frequencies, and thus the cutoff
layers are located at different major radii. The power spectra from
all the first three channels exhibit clear exponential shapes. Figure~\ref{spectra_R}
also shows a decreasing trend in $\tau$ as the major radii of the
cutoff layer decreases.  At the innermost radial location, $r/a =
0.82$, while the spectrum is approximately consistent with an exponential, the low signal-to-noise ratio makes it harder
to determine the shape.

\subsection{Lorentzian pulses}
Pulses with an approximately Lorentzian shape are found in the time signals from the reflectometer,
consistent with the observation of exponential power spectra \cite{winters2014deterministic}. A fitting routine was developed to locate and fit Lorentzian pulses in the time history. The routine takes a small time window of $n$ points with length $n\Delta t,\ 7 \leq n \leq 14$. It finds the extrema $x_0$ in the chosen window at time $t_0$, and fits the points in the time window to a Lorentzian function that passes through $(t_0,\ x_0)$.

\begin{figure}[h]
\centering
\includegraphics[width = \linewidth]{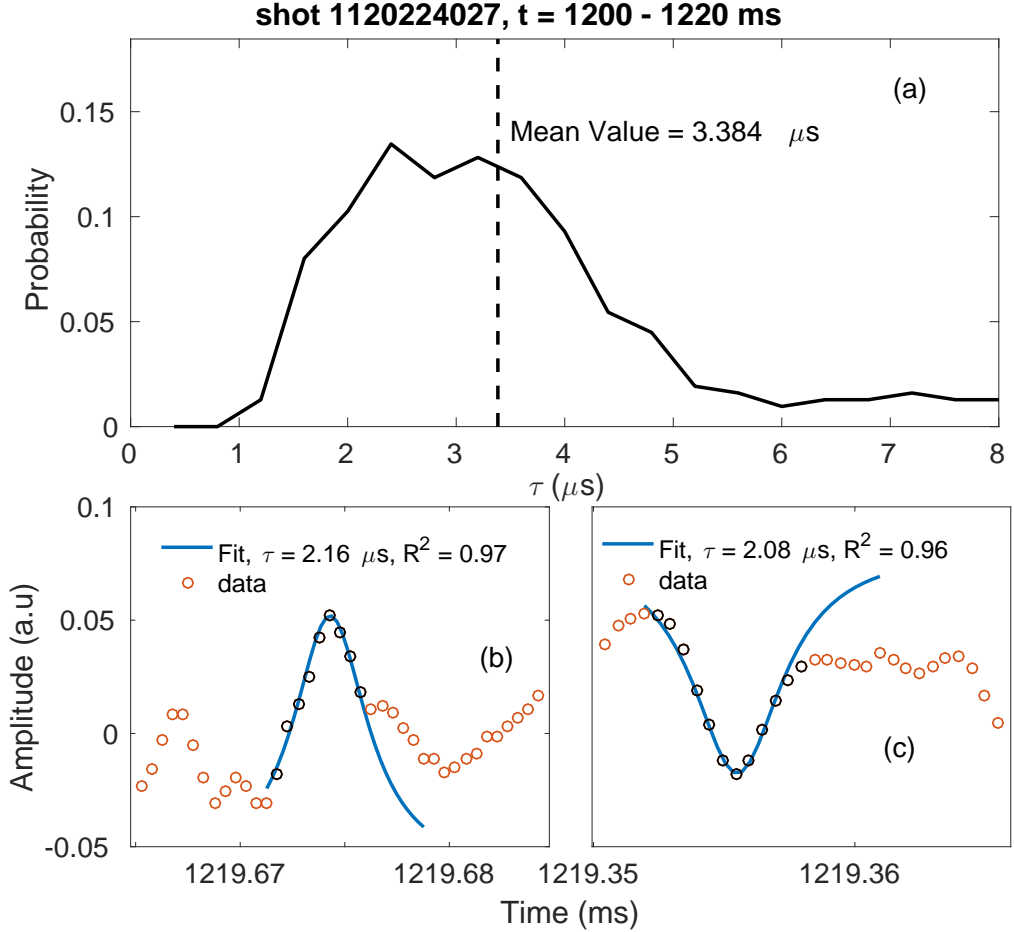}
\caption{(a) shows that the distribution of mean value of pulse widths found in the time history is larger than the $\tau$ calculated by fitting the power spectrum (fig. \ref{spectra} and fig. \ref{spectra_R}). (b) and (c) are two typical Lorentzian pulses found by the automatic fitting routine; only the peak of pulses are visible due to superpositions. These data are from the reflectometer time series signals.} 
\label{dist}
\end{figure}

\begin{equation}
L(t) = A\ \left[\frac{1}{\tau^2 - (t-t_0)^2} - \frac{1}{\tau^2}\right] + x_0
\end{equation}
where A is a normalization constant and $\tau$ is the pulse width of the pulse, defined to be half width at half maximum of the pulse.

If the fitted curve and the data agree within 5\% on every point, the
points in this time window are documented to be a pulse and then the
routine moves to the next time window. When finding pulses, the
routine choses to start with larger $n$ such that the tails of a pulse
is best captured while also avoiding counting pulses repeatedly.

The $R^2$, coefficient of determination, quantifies how well a curve
fits to the data is calculated for fits to a Lorentzian function. For a given dataset $\{y_i\}$ and fit $\{ f_i
\}$, $R^2 = 1-\frac{SS_{\mathrm{err}}}{SS_\mathrm{tot}}$, where $
SS_\mathrm{err} = \sum{(y_i - f_i)^2}, SS_\mathrm{tot} = \sum{(y_i - \bar{y})^2}$.

\begin{figure}[h]
\centering
\includegraphics[width = \linewidth]{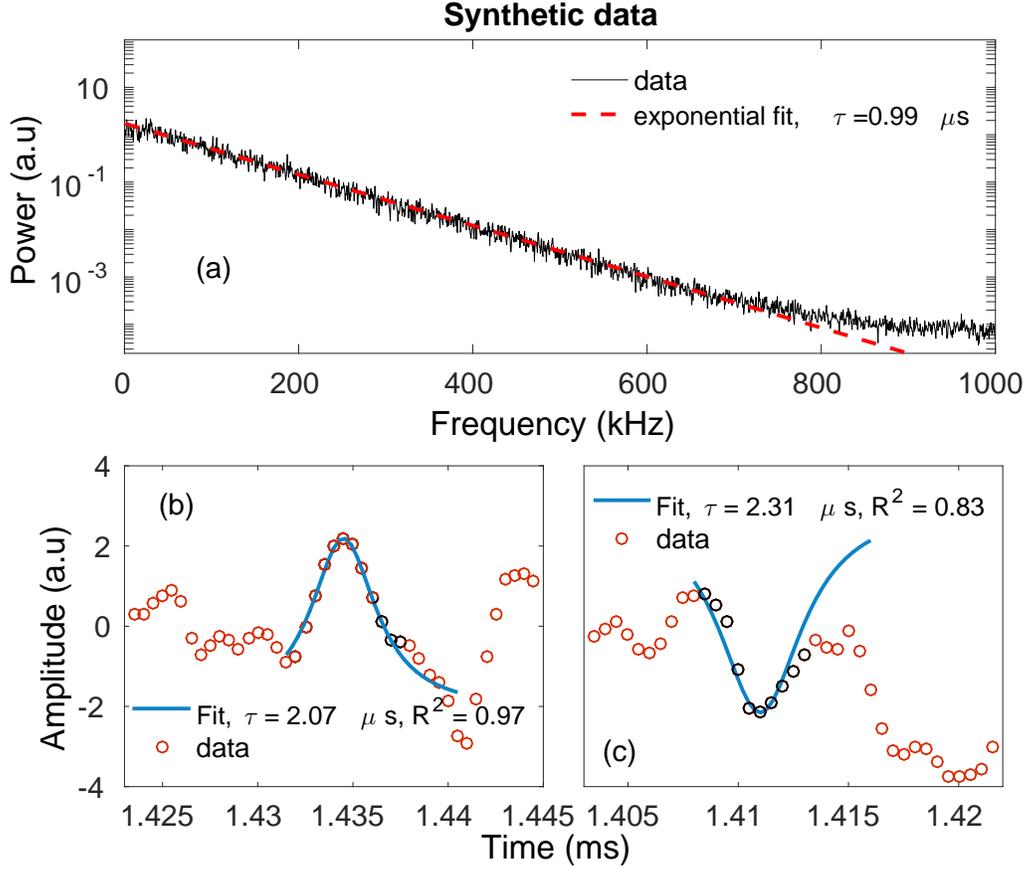}
\caption{Synthetic data of overlapping pulses with pulse width $1 \mu s$. The pulse width calculated by fitting the power spectrum shows an agreement with the actual pulse width in (a). The pulse widths found by fitting time history using the automatic routine in (b) and (c) are about twice of the actual value due to superposition.}
\label{overlapping}
\end{figure}

Figure~\ref{dist}(a) shows a distribution of pulse widths found by the fitting routine. The average pulse width and pulse width
found by fitting the power spectrum do not always agree. This disagreement could be explained by the overlap of pulses in the temporal signal (Fig.~\ref{dist}(b) and (c)). Simulations of overlapping pulses have indicated that overlapping pulses can cause the pulse widths determined by fitting the power spectrum to be greater than the actual value determined by fitting the time signal (Fig. \ref{overlapping}). However,
the value of $\tau$ found by fitting the power spectra reflect the input pulse width(Fig. \ref{overlapping}(a)). Therefore, fitting power spectra is a more sensitive method and it is used to find the pulse width for the rest of this work.

\begin{figure}[h]
\centering
\includegraphics[width = \linewidth]{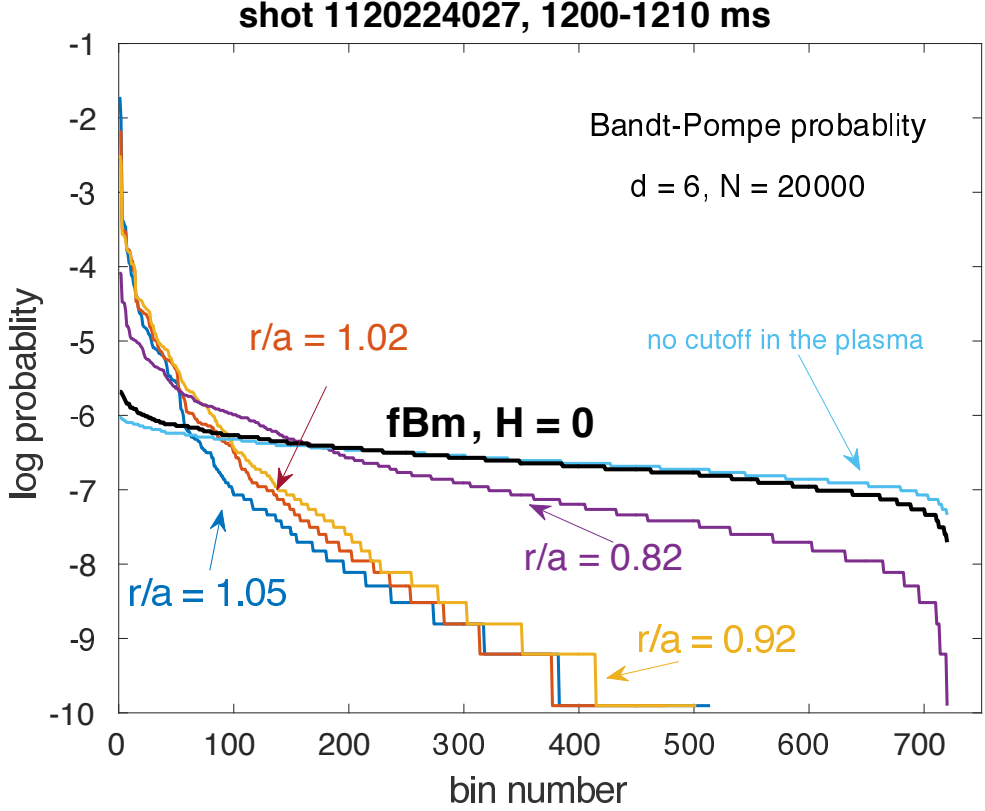}
\caption{The Bandt-Pompe probability distributions corresponding to the power spectra in figure \ref{spectra_R} shows edge fluctuations have preferable states while core fluctuations do not.}
\label{BP_radii_Lmode}
\end{figure}

\begin{figure}[h]
\centering
\includegraphics[width = \linewidth]{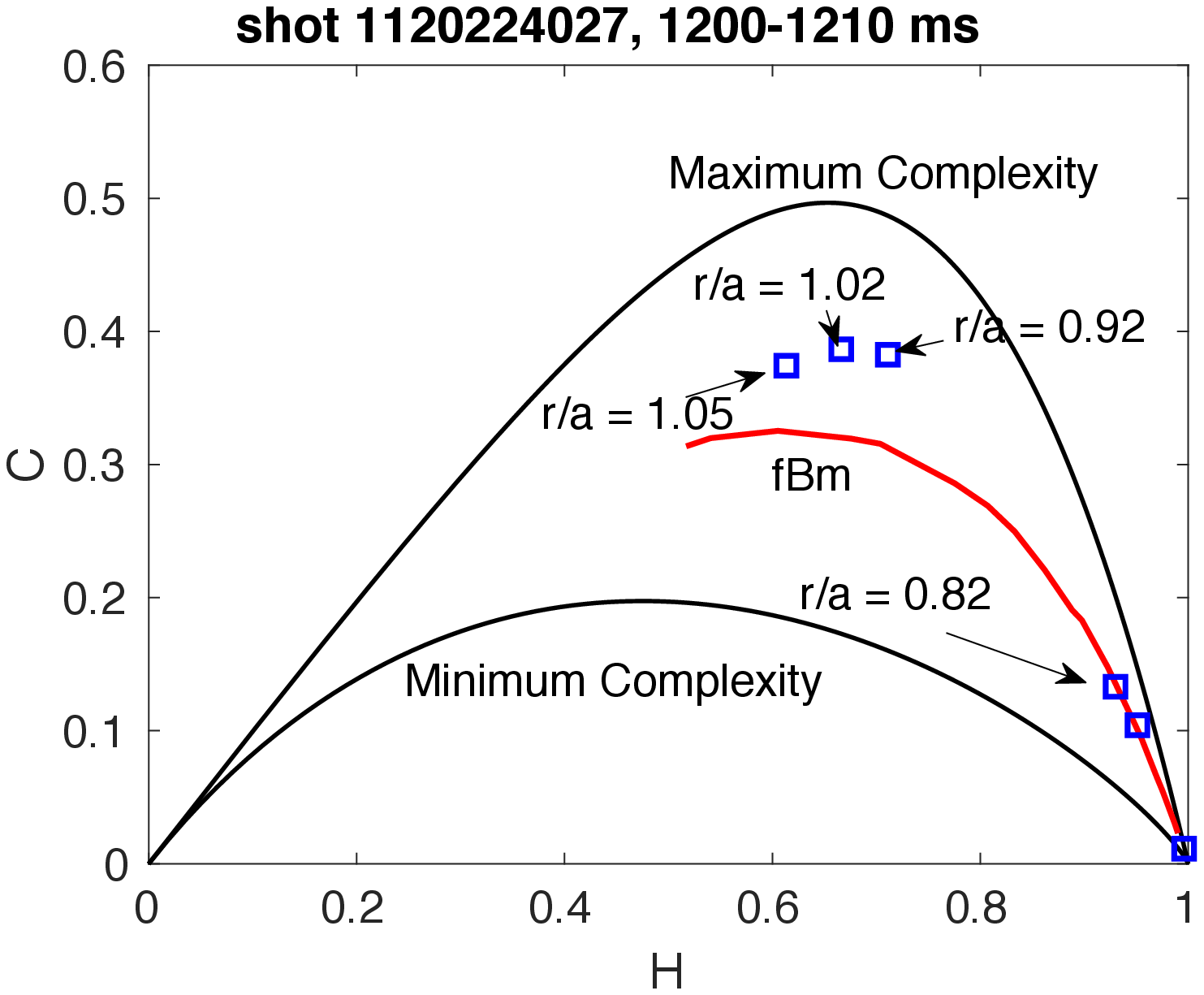}
\caption{The C-H plane locations corresponding to the BP probability distributions shown in figure \ref{BP_radii_Lmode} indicate that edge turbulence is strongly chaotic.}
\label{CH_radii_Lmode}
\end{figure}

\subsection{BP probability and C-H plane}

 Figure~\ref{BP_radii_Lmode} shows the BP probability
distributions of the inphase signal corresponding to the same shot in
Fig.~\ref{spectra_R}. The BP probability distribution for measurements
at $r/a = 1.05,
1.02$ and $0.92$ are qualitatively similar to the distribution for the
double pendulum: there is a clear structure in the signal, with
particular amplitude ordering permutations dominating the signal.  At $r/a =
0.82$, the BP distribution looks more similar to noise (near uniform
distribution across permutations); this is consistent with the low
signal-to-noise at this location, so that the stochastic noise signal
dominates the BP distribution.  One additional BP distribution is
shown in this figure for the case where the frequency of the
reflectometer is too high to reflect within the plasma.  This signal
should primarily contain noise (electronic) and is consistent with
fBm. 

These BP probability distributions are used to place the time signals
on the C-H plane, which is showin in Fig.~\ref{CH_radii_Lmode}.
The fluctuations in the edge region ($r/a = 1.05, 1.02, 0.92$) are
clearly in the chaotic region of the CH plane, with high complexity
and moderate entropy.   The data at $r/a \sim 0.86$ is located near
the fBm line,  which shows that the signals measured in the core are
more stochastic.  This is consistent with the low signal to noise at
this location, where the dominance of electronic noise could explain
the placement on the C-H plane. 

\subsection{Gas puff imaging (GPI) measurements}

\begin{figure}[h]
\centering
\includegraphics[width = \linewidth]{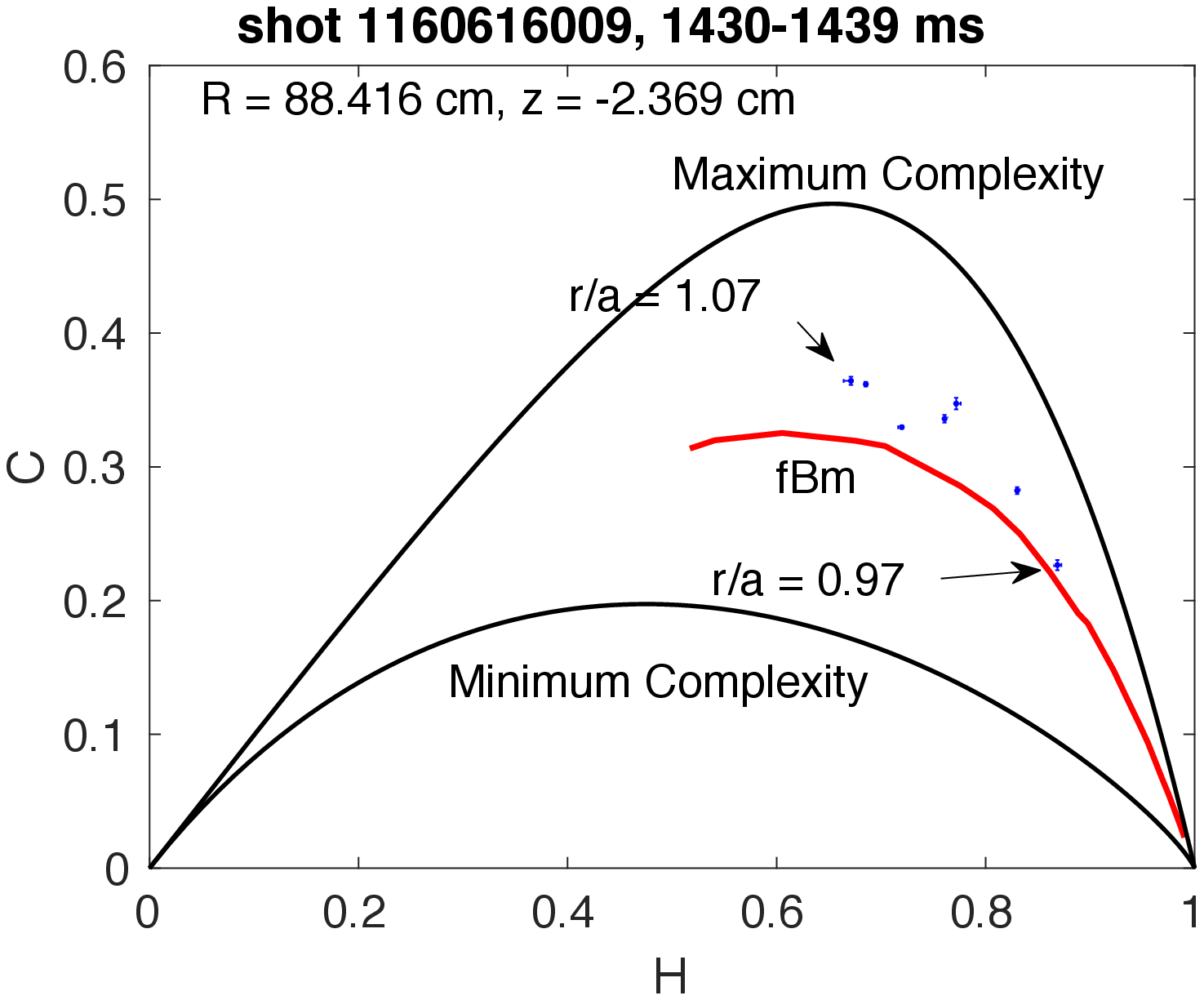}
\caption{The C-H plane locations of GPI show edge fluctuations are chaotic. Errorbars come from subsampling the signals.}
\label{CH_GPI}
\end{figure}

\begin{figure}[h]
\centering
\includegraphics[width = \linewidth]{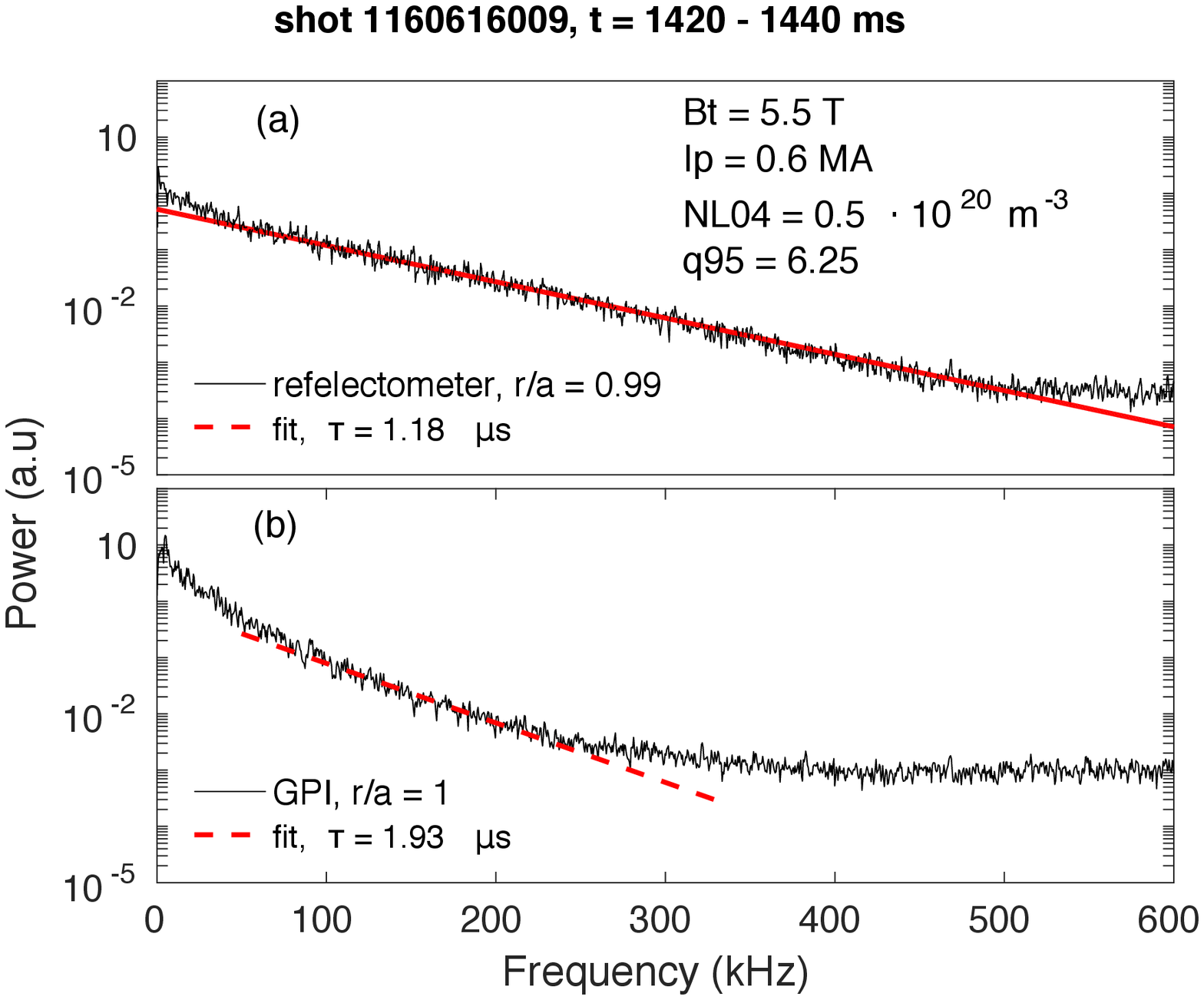}
\caption{A GPI spectrum is compared with the reflectometer spectrum from the same shot and same time interval. Both spectra have exponential shapes but they have different slopes. }
\label{spectra_GPI}
\end{figure}

The same analyses (power spectra, BP probability, the C-H plane) were
performed on the GPI data. Figure~\ref{spectra_GPI} compares a
reflectometry inphase spectrum and a GPI spectrum at roughly the same
radial position (r/a $\sim$ 0.99). Both spectra show an approximate linear shape on
semi-log scales that is consistent with exponential power spectra.  While
reflectometry and GPI show the same result qualitatively, the GPI
spectrum has a larger slope on this scale, which corresponds to a
larger pulse width in time. The reason for this is presently unknown. Possible explanations include the fact that the exact cutoff layer of the
reflectometry is undetermined; the radial position measured by the
reflectometer might be slightly smaller than that from the
GPI. Figure~\ref{spectra_R} shows that the spectra with smaller major
radii have smaller pulse widths, which is a possible explanation of
the discrepancy in slopes. Furthermore, as mentioned in Section II, GPI-measured fluctuations depend on both density and
temperature fluctuations and can be approximated as $\tilde{S}/\bar S \approx \alpha \tilde{n_e}/\bar n_e + \beta \tilde{T_e}/\bar T_e$, with $\alpha \sim$ 0.5 and $\beta \sim$ 0.1 under the conditions for this shot \cite{cziegler2011turbulence}. Depending on the amplitude of $\alpha \tilde{T_e}$ may or may
not have a significant contribution to $\tilde{S}$. Therefore, the GPI
was measuring a somewhat different quantity than the reflectometry.

Figure~\ref{CH_GPI} shows the C-H plane locations of GPI fluctuation
data at different major radii. GPI data at different radii require
different subsampling rates as GPI has a higher noise floor than
reflectometery (fig. \ref{spectra_GPI}). The subsampling rates chosen
for the 7 radial positions are $[5, 5, 8, 3, 3, 3, 3]$
respectively. They were chosen based on the noise level and the C-H
locations of different sampling rates. The C-H locations of different
radii follow the same trend as reflectometry: the signals are more
chaotic in the SOL and have a more dominant stochastic component closer to the core.
Therefore, both power spectra and the C-H plane analysis from GPI data
support the chaotic nature of the C-Mod edge density fluctuations.

\subsection{Overview of H-mode and I-mode edge density fluctuations}
\FloatBarrier

\begin{figure}[h]
\centering
\includegraphics[width=\linewidth]{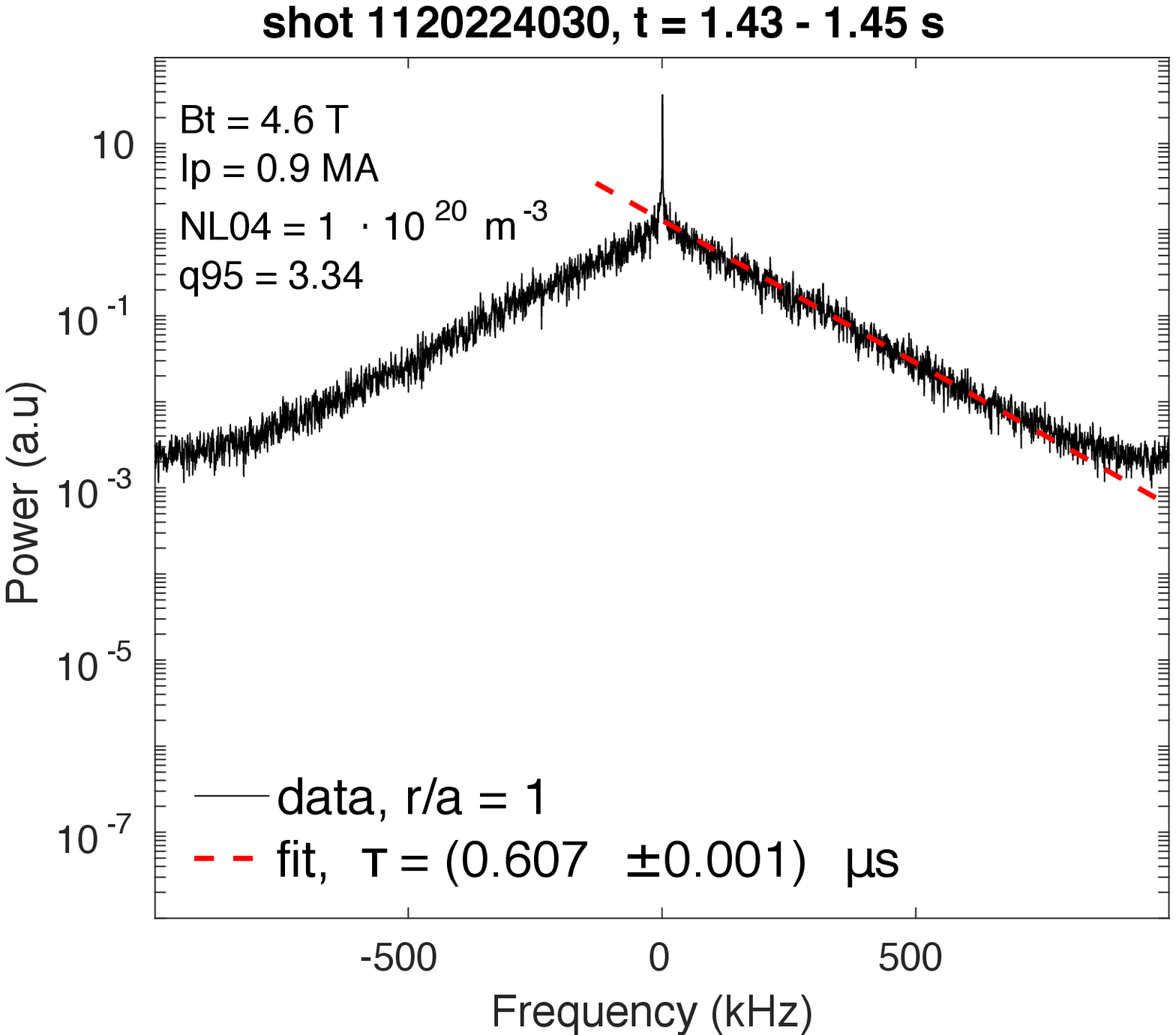}
\caption{A power spectrum of fluctuations measured with the reflectometer for an H-mode plasma near the SOL exhibit an exponential shape.}
\label{spectra_H}
\end{figure}

\begin{figure}[h]
\centering
\includegraphics[width = \linewidth]{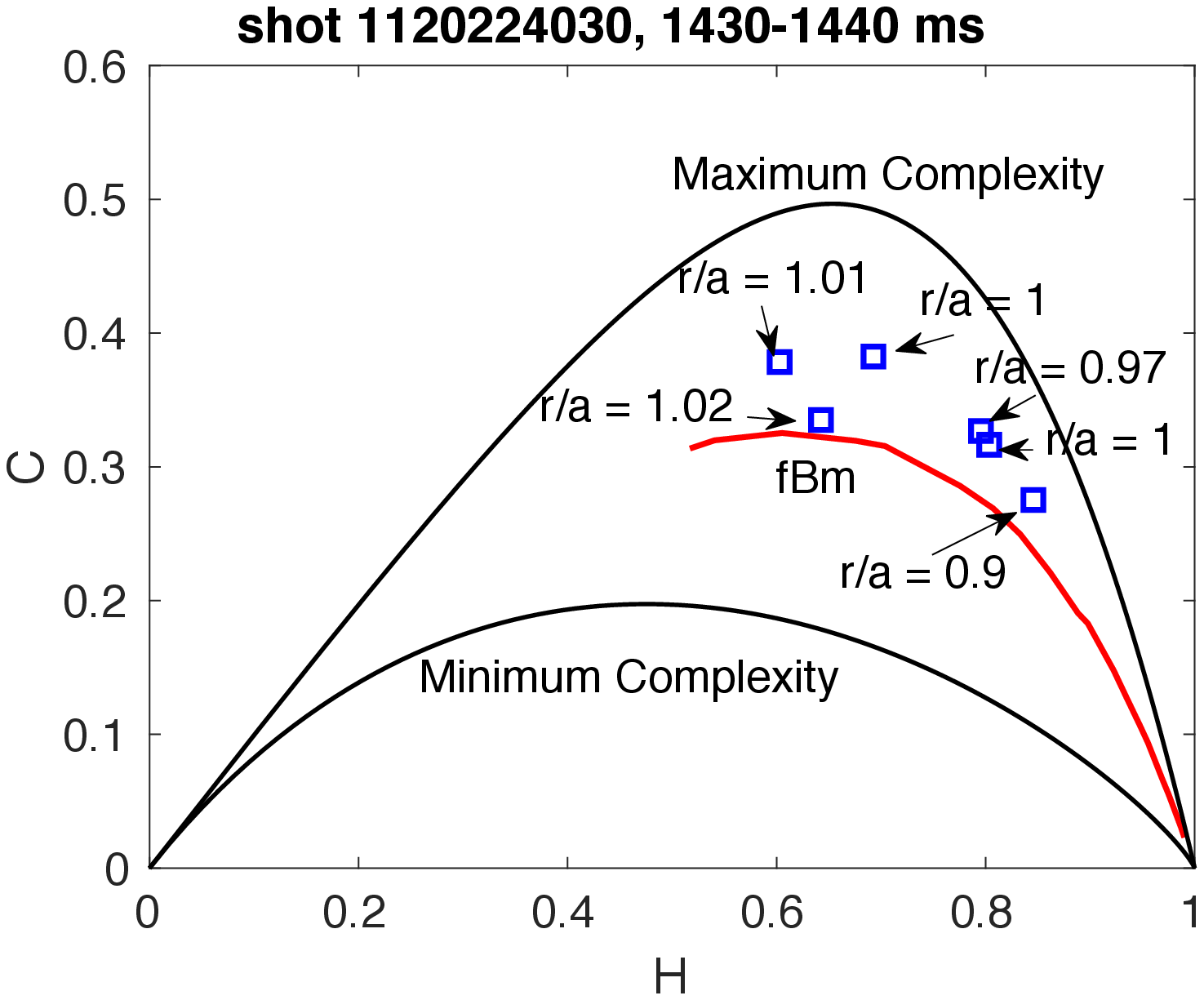}
\caption{The C-H plane locations corresponding to figure \ref{spectra_H} (H-mode) indicate the chaotic nature of H-mode edge density fluctuations.}
\label{CH_Hmode}
\end{figure}

\begin{figure}[h]
\centering
\includegraphics[width = \linewidth]{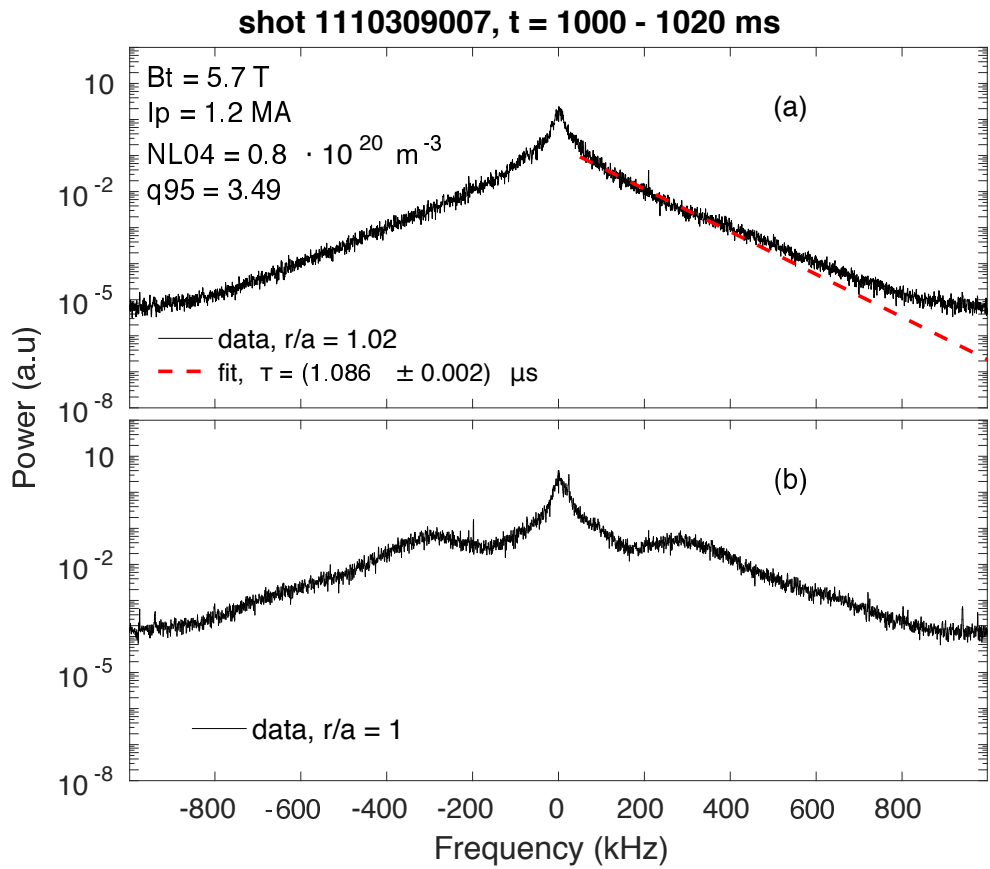}
\caption{Power spectra of reflectometry-measured fluctuations from an I-mode plasma (a) outside and (b) just inside SOL are compared. The spectrum outside the SOL exhibits exponential shape. The weakly coherent mode (WCM) appears around 200 kHz in the power spectra measured inside or at the SOL, making fitting the spectrum difficult.} 
\label{spectra_I}
\end{figure}

\begin{figure}[h]
\centering
\includegraphics[width = \linewidth]{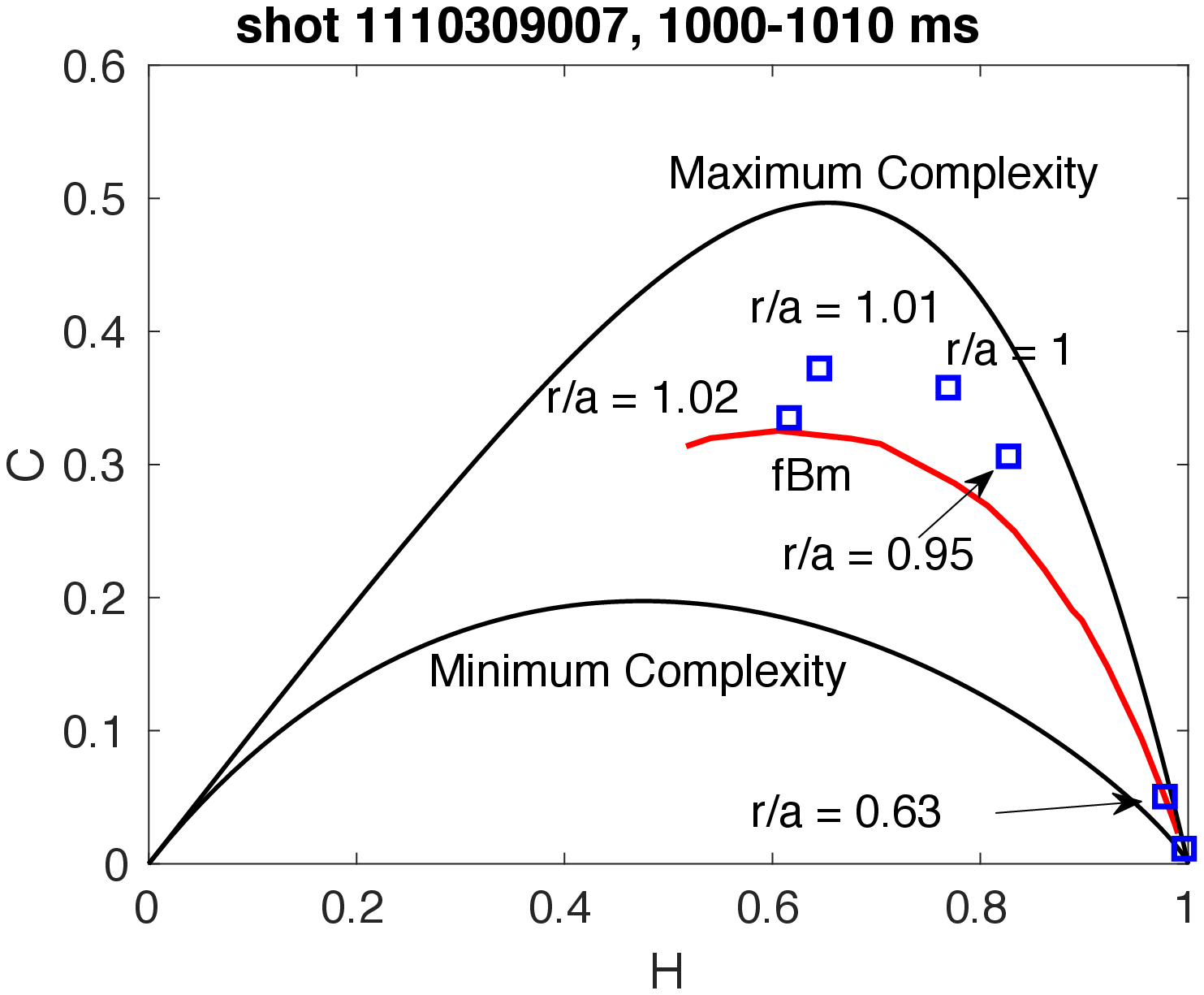}
\caption{The C-H plane locations corresponding to figure \ref{spectra_I} (I-mode) indicate the chaotic nature of I-mode edge density fluctuations.}
\label{CH_Imode}
\end{figure}

In addition to L-mode plasmas, H-mode and I-mode plasmas were
studied. H-mode plasmas are characterized by sharp edge density and
temperature gradients~\cite{Connor2000}. I-mode plasmas have an edge
temperature gradient but their density profile is almost identical to
L-mode; they have the same thermal-transport as H-mode but similar
particle transport as in L-mode. ~\cite{Whyte2010}

Figure~\ref{spectra_H} shows a power spectrum of
reflectometry-measured edge fluctuations measured during an H-mode
plasma, which exhibit an exponential shape. Similar to L-mode
plasmas, pulses were identified in the time signal in this
case. Figure~\ref{CH_Hmode} shows the C-H plane locations corresponding to
the power spectra in Fig.~\ref{spectra_H}. The location on the C-H
plane and trend with changing radius in this case is similar to what
was observed in L-mode plasmas (see Fig.~\ref{CH_radii_Lmode}).  The shape of the power spectrum and the
C-H plane analysis indicate that H-mode edge fluctuations are generated
by a chaotic process. 

Figure~\ref{spectra_I} shows power spectra of reflectometry-measured edge fluctuations during an I-mode
plasma, one just into the SOL ($r/a = 1.02$) and one just at the
separatrix. The peak near 200 kHz in the spectrum measured at the
separatrix is the weakly coherent mode (WCM),
which is a typical feature of I-mode edge power
spectra. The WCM is radially localized, and it does not
appear in the SOL. Pulses are observed in the 
$r/a \sim 1$ time signal. However, the appearance of WCM makes it hard
to determine if the remainder of the broadband spectrum is consistent
with an exponential model. Therefore, the C-H analysis is necessary to
identify the nature of the fluctuations with the presence of
WCM. More extensive study of the WCM properties, such as interaction with the Geodesic Acoustic Mode (GAM), is beyond the
  scope of this work, but can be found in {\it Cziegler} \cite{cziegler2013fluctuating}.
The C-H plane locations of
the I-mode plasma support the conclusion that edge fluctuations are
produced by a chaotic process. Figure \ref{CH_Imode} shows a similar trend as the L-mode
plasmas: the signals are strongly chaotic on the edge and have large
noise component as measurements move toward the core.  

\section{Discussion and Conclusions}~\label{dis}
\FloatBarrier 

Edge density fluctuations measured in the L-mode, H-mode and I-mode plasmas using reflectometry and GPI are
found to be chaotic. The reflectometry-measured signals have exponential power
spectra and, consistent with the spectral shape, contain
Lorentzian-shaped temporal pulses. Note that the GPI-measured spectra have a slight different shape, which might be due to the signals' dependence on temperature fluctuations. In addition, the complexity and
entropy of the measured signals are computed and shown to fall on the chaotic
region of the C-H plane. The exponential power spectra and the C-H locations together
support the chaotic nature of the edge density fluctuations. The core
turbulence is more stochastic on the C-H plane. However, with the
small signal to noise ratio from the reflectometer in the core region,
it is hard to establish the real nature of core turbulence, and
new experiments will be required.

Similar experiments and analysis to those presented here
were performed previously on many other toroidal and linear devices
using different diagnostic techniques: the DIII-D tokamak
\cite{Maggs2015} with the Doppler backscattering (DBS), the TJ-K
stellerator,\cite{Hornung2011} and the Large Plasma Device (LAPD)
\cite{PacePRL, Carter2006} with Langmuir probes. 
In these experiments, Lorentzian-shaped pulses and exponential power
spectra were identified and the chaotic nature of the edge density
fluctuations was established. The normalized time scale associated
with the exponential power spectra, $\tau f_{\mathrm{ci}}$, are within a factor of 2 in all these cases despite a wide
  range of plasma parameters.~\cite{Maggs2015} For
example, the typical magnetic field in the LAPD is 1000 G and that of
the DIII-D tokamak is 3 T. The Langmuir probe data in the TJ-K yield
the range of $\tau f_{\mathrm{ci}}$ between 3.7 and 5; experiments in
the LAPD\cite{PacePRL, Carter2006}
reveal $\tau f_{\mathrm{ci}} = 6$ and  in the DIII-D Doppler
backscattering data, $\tau f_{\mathrm{ci}}$ = 8.6 on average. However,
the time scale in the Alcator C-Mod tokamak is drastically different
than all of these other cases, with $\tau
f_{\mathrm{ci}}$ around 30 - 50.  Parameters of different devices are compared
in table \ref{param}, and it is not obvious which parameter controls the scaling of $\tau f_{\mathrm{ci}}$. A theoretical
explanation of the large value of $\tau f_{\mathrm{ci}}$ compared to
DIII-D \cite{Maggs2015} and LAPD and its relation to plasma transport models could be valuable to develop.  

\begingroup
\squeezetable
\begin{table}[h]
\caption{\label{param} Comparison of parameters between different experiments; $\nu_{e}$ denotes the electron collision rate, $\nu_{e} = 1.33\cdot10^5 n_e T_e^{-3/2}$.}
\begin{center}
\begin{ruledtabular}
\begin{tabular}{c  c c c c}

Device  &   $ n_e (10^{20}\cdot \mathrm{m^{-3}}) $ &$ T_e\ (\mathrm{keV}) $ & $\nu_{e}\ (10^5 \cdot \mathrm{s^{-1}})$ & $\tau f_{ci}$\\

C-Mod (L-mode) & 0.35 & 0.1 & 14.7 & 40\\
C-Mod (H-mode) & 1 & 0.3 & 8.1 & 27\\
DIII-D & 0.26 & 2 & 0.12 & 8.6\\
LAPD & 0.01 & $ 5\cdot 10^{-3}$ & 37 & 4\\
TJ-K & $10^{-3}$ & $7\cdot 10^{-3}$ & 2.3 & 3.7-5 \\
\end{tabular}
\end{ruledtabular}
\end{center}
\end{table}
\endgroup

The results from the
Alcator C-Mod tokamak along with other toroidal and linear devices suggest that fluctuations in the edge region of magnetically-confined
plasmas are generated via a chaotic process.  
Establishing the chaotic nature of edge turbulence in these devices
provides a guide for further work the understand and perhaps control the processes that
lead to turbulence and transport.  In systems exhibiting low-dimensional chaos,
as few as two interacting linear modes can explain the dynamics.
Controlling one of these modes using, e.g. nonlinear three-wave
interactions~\cite{auerbach2011resonant}, may lead to the ability to control edge
transport. 

For future work, we note that this study did not establish the nature of the core density fluctuations. Although 
  the analysis suggests that turbulence measured in the core region is
  chaotic, with large noise components, whether or not core turbulence
  is dominantly stochastic or chaotic in nature cannot be determined
  with the current set diagnostics considered here. Additional and new
  experiments on other tokamaks should be carried
  out. Exploring the chaotic nature of core turbulence
  could also be done theoretically, using, for example, nonlinear
  gyrokinetic simulations \cite{garbet2010gyrokinetic} to generate
  long-time series and then performing the same analysis as done here
  for experimental time series.
  
  The diagnostics in this work could be improved. While reflectometry
does not require time-averaging and does not perturb the plasmas, it has variation in its measured positions. Several effects, including 2D effects and scattering, could also impact the
  interpretation of the results. GPI measures fluctuations at a relatively well-fixed position in space on a 2-D grid but had higher noise floor and might be influenced by temperature fluctuations, which makes direct comparisons between GPI
  and reflectometer difficult. Meanwhile, the injected
neutral gas used for GPI could potentially perturb the
plasma and complicate interpretation. Therefore, additional detailed modeling of the reflectometer and GPI
  diagnostics is required. Related to this, it could be very useful to
  generate either ``toy-model'' time series of the turbulence or use
  turbulence simulation time series outputs, and apply ``synthetic
  diagnostics" that mimic the reflectometer and GPI \cite{holland2009implementation}
  signals and look for evidence of chaotic dynamics in these synthetic
  time series. It would also be beneficial to use a
  third diagnostic, such as the mirror Langmuir probe at C-Mod,\cite{labombard2007mirror} to compare with reflectometry and GPI data.

Some studies of C-Mod GPI data, as well as Langmuir probe data from K-STAR and TCV indicate stochasticity in edge fluctuations.~\cite{garcia2013intermittent, garcia2013burst, kube2016fluctuation, theodorsen2016scrape, garcia2016sol} Recent modeling work has shown that a random distribution of pulse widths can result in power law spectra, even if the model time series is composed of only Lorentzian pulses.~\cite{garcia2016power}  However, previous works~\cite{garcia2013intermittent, garcia2013burst, kube2016fluctuation, theodorsen2016scrape, garcia2016sol} did not include the C-H plane analysis, and we have demonstrated that applying the C-H analysis to a set of synthetic data with Lorentzian pulses with randomly distributed pulse widths will indeed show its stochastic nature (the blue triangle in fig. \ref{fig:CH_test}). Therefore, the C-H plane analysis is capable of distinguishing the chaos and stochasticity of this type of synthetic data, and a chaotic description is valid for the data presented in this study. This suggests that future studies should include C-H plane analysis as a technique to show the nature of turbulence fluctuation data. 
  
While the C-H plane analysis can be used as evidence to support the conclusion that the fluctuations are chaotic, it has to be combined with other methods such as power spectra analysis, because there is no definite boundary between chaotic and stochastic signals.  It is also difficult to compare the C-H locations of different reflectometer channels because there is no parameter that accesses the ``degree of chaos" of a signal. Therefore, developing such a parameter to quantify
the ``degree of chaos'' can be beneficial for understanding the turbulent
nature. Using the power spectrum as a part of a test for chaotic
  dynamics though also involves some uncertainty. For example, real
  diagnostic data often do not always have a wide enough dynamic range
  (e.g. a low noise floor) to capture enough orders of magnitude in
  the measured spectrum to faithfully capture the shape. Therefore, all three techniques, power spectra, the BP probability distribution, and the C-H plane, have to be combined to identify the nature of turbulent fluctuations, and it would be valuable to apply these techniques to analyze the data in other devices with various diagnostics.

\acknowledgements{
Discussions with G. Morales, J. Maggs, J. Bonde, M. Martin, and S. Dorfman are gratefully acknowledged. The authors thank V. Winters for her previous work. The experiments reported in the paper were performed in the MIT Plasma and Fusion Center. This work was supported by the U.S. Department of Energy Office of Science under Agreement DE-FC02-99ER54512 and DE-FC02-07ER54918-011.}

\section*{References}
\bibliography{mybib}

\end{document}